# Enhanced Strain Transfer and Optoelectronic Performance in MoS$_2$ Devices via Formvar Encapsulation

*Simeon N. Vladimirov,[1] Onur Çakıroğlu,[1] Carmen Munuera,[1] Andres Castellanos-Gomez,[1] Thiago L. Vasconcelos [1,+]*

AUTHORS ADDRESS

[1] 2D Foundry research group, Instituto de Ciencia de Materiales de Madrid (ICMM-CSIC), Madrid E-28049, Spain.

[+] Present address: Materials Metrology Division, Instituto Nacional de Metrologia Qualidade e Tecnologia (INMETRO), 25250-020 Duque de Caxias, RJ, Brazil

**ABSTRACT**

We systematically investigate the influence of polyvinyl formal (PVFM), commonly known as Formvar, in comparison to polycarbonate (PC) and polymethyl methacrylate (PMMA), as encapsulation materials on the strain performance of $MoS_2$ monolayer and bilayer flakes on flexible polypropylene (PP) substrates. Notably, optical differential reflectance measurements reveal that PVFM and PMMA encapsulation significantly enhances the mechanical and thermal strain gauge factors by approximately 2-fold (up to ~-50 meV/%) and 6-fold (up to ~-1.5 meV/°C), respectively, while PC shows a slightly lower enhancement. Moreover, all three polymers increase the maximum achievable strain from approximately 1.4% to 2.3%. Furthermore, devices fabricated on PP substrates exhibit improved optoelectronic performance when encapsulated with PVFM, including increased and faster photocurrent response and extended device lifetime.

1. **INTRODUCTION**

The capability to change and control the optical and electrical properties of two-dimensional (2D) semiconductors by an external tuning knob is a key element in the advancement of modern electronics.[1] In particular, it is very appealing that their properties can be modified by deforming their crystal lattice in what is called strain engineering. Strain alters the lattice constants, inducing changes in the electronic band structure and leading to shifts in the excitonic features. Compressive strain reduces the interatomic distances, increasing the overlap of atomic orbitals, which can narrow the band gap. Tensile strain, on the other hand, increases the interatomic distances, decreasing orbital overlap, which can widen the band gap.[2] This strain engineering is a powerful tool, especially for 2D materials as they can be easily stretched and bent to a relatively large extent in a reversible way.[3,4] For example, by applying mechanical strain to a few atomic layers of transition metal dichalcogenides (TMDCs), their energy bandgap can be controllably reduced by tens of meV.[2,5,6] Moreover, the encapsulation of the 2D material with specific polymers can even improve the gauge factor,[7,8] defined as the ratio of the bandgap energy shift to the percentage change in strain, while at the same time it can protect the device leading to a longer device lifetime and improve its optical response time and stability.[9]

Encapsulation strategies have been widely employed across various systems using different materials to prevent ambient degradation[10–13] Specifically, polymer capping has emerged as an efficient and scalable approach for protecting 2D materials from environmental exposure. This is particularly critical for TMDCs, Especially for the TMDCs where the chalcogen vacancies on the surface interact with reactants of the ambient air, leading to strong degradation in a relative short time, even in the range of few hours.[9] Capping with different polymers, including PMMA,[14–17] polyvinyl pyrrolidone (PVP),[18] and polyvinyl alcohol (PVA)[19] has been showed to improve stability of 2D materials in air. Among them, PMMA is one of the most used polymers, leading to highly stable electron and hole mobility in FETs based on TMDCs even after several weeks.[15]

It has been also proven that this encapsulation improves the strain transfer and gauge factor. The mechanism behind the strain transfer improvement by applying a polymeric encapsulation was reported to be attributed to two main components. First, the improvement of the Young's modulus for the substrate-2D material-polymer encapsulation assembly plays a crucial role. Flexible polymer substrates typically used in strain devices have a much lower Young's modulus compared to 2D material samples. Additionally, the weak van der Waals forces between the substrate and the 2D material often result in incomplete strain transfer from the substrate to the 2D semiconductor[6,8,20] By capping with a polymer that has a higher Young's modulus and a stronger interaction force with the 2D material, the strain transfer can be significantly enhanced.[7,8,21] Second one is that with the encapsulation, the material and substrate are bound together by polymer grabbing forces reducing the chances of slippage of the flake and expanding the applicable strain range. Therefore, the maximum strain that can be applied before the 2D material slips or breaks, which is another important figure of merit in straintronics, can be extended.[21]

Experimentally, improved strain gauge factors and maximum achievable strain were observed for polymers capping, including PVA,[8] polydimethylsiloxane (PDMS),[22] PMMA,[23,24] and a staking capping made of Graphene/Formvar(PVFM)/PDMS.[21] Recently, our group has demonstrated an unprecedented crystal strain of 2.8% and maximum strain gauge factors of -99.5 meV/% and -63.5 meV/% for the A and B exciton of monolayer $MoS_2$ sample by covering it with an adamantane ($C_{10}H_6$) plasma polymer pinning layer.[7] However, the adamantane cover layer is produced by remote plasma-assisted vacuum deposition, which is not easily accessible.

In this work, we present a systematic study of the influence of three transparent, easily accessible polymers on tensile strain transfer and maximum achievable strain in $MoS_2$ monolayers and bilayers, as assessed through optical differential reflectance measurements. The encapsulation method employed is quite simple, involving spin coating of solutions prepared by dissolving the polymers in a solvent or using polymers purchased in a ready-made solution. Notably, we demonstrate that the encapsulation using PVFM (Formvar), a polymer known for its excellent electrical insulation and high resistance to radiation damage,[25] significantly enhances both strain transfer and device performance. This simple yet effective encapsulation method not only improves strain gauge factors but also leads to remarkable enhancements in photocurrent response, with increased response speed and extended device lifetimes. Additionally, strain-induced energy shifts were mapped through photoluminescence (PL) hyperspectral imaging and also assessed via spectral response measurements of photosensors without and with PVFM encapsulation.

## 2. METHODS

MoS$_2$ monolayer or bilayer flakes (mineral Molybdenite from Molly Hill Mine – LaMotte Township, Quebec, Canada) were exfoliated onto Polydimethylsiloxane (PDMS) Gel-Film substrate (WFx4 6.0 mil, by Gel-Pak) and transferred to a 500 μm thick PP substrate (PP30-FM-000340, Goodfellow Cambridge Limited) as described in Ref.[26] Differential micro-reflectance spectra were used to identify the number of layers of MoS$_2$ on the selected samples.[27] This sample preparation steps mimics what is conventionally done for the fabrication of TMDC-based devices on polymer substrates and the results can be better generalized. The flexible PP substrate was chosen for its high thermal expansion coefficient (in-plane $73 \times 10^{-6}$ K$^{-1}$, out-of-plane $118 \times 10^{-6}$ K$^{-1}$),[28] high Young's modulus (~3.5 GPa),[28] and high resistance to many organic polymer's solvents. All PP substrates used in this study were previously flattened by placing them between two Si wafers at 145 °C for 6 min. During this process, a weight was applied on top, exerting approximately 13 kPa of pressure to ensure uniform flatness.

A total of 26 MoS$_2$ samples were fabricated, each with an area exceeding 40 μm$^2$. Measurements were taken at least 3 μm from the edge to minimize distortion due to edge effects.[29] The sample set comprised 14 monolayers and 12 bilayers, yielding 30 mechanical strain measurements and 28 thermal strain measurements. Among these, 11 mechanical strain measurements (8 for thermal strain) were conducted on unencapsulated samples, while 7 samples were encapsulated with PC, 6 with PVFM, and 6 with PMMA. Detailed results for both mechanical and thermal strain measurements are provided in the Supporting Information. The polymer encapsulation was done by spin coating (3000 rpm and 40 s) 50 μL of the dissolved polymer solution on the top of the sample. The polymers solutions tested were PC (~1 % w/w in Acetone), PMMA (~1.2 % w/w in 1,2-dichloroethane) and PVFM (~1 % w/w in 1,2-dichloroethane, Formvar solution from Merck company). The polymers solutions are easy to prepare and can be done by simply dissolving the solid polymers in 1,2-dichloroethane or Acetone using a magnetic stirrer for 12 hours. This method leads to a (110 ± 50) nm thick capping polymer film (Figure S1 on Supplementary Material).

The absorbance spectra were evaluated as the tensile strain was applied first by thermal expansion and later by mechanical uniaxial strain using a three-point bending motorized setup in the same way as shown on Ref.[5] For the PL measurement, one $MoS_2$ monolayer sample was transferred to PP and encapsulated with PVFM. The PL hyperspectral images were carried out with a MonoVista CRS+ system (Spectroscopy and Imaging GmbH) with 532 nm laser excitation using a 100× objective with a laser power of 0.2 mW and an integration time of 3 seconds per pixel of 1 μm size. Diffraction gratings of 300 lines/mm were used. They were acquired at 0% and 1% mechanical uniaxial strain achieved with the aforementioned setup. All spectral fitting and PL hyperspectral map adjustment fit was performed using FabNS's PortoFlow Analysis Software.

For the device fabrication, aimed at evaluating the impact of polymer encapsulation on photocurrent response, several $MoS_2$ flakes were first exfoliated from a bulk crystal onto a PDMS sheet and then deterministically transferred onto the top of electrode contacts on PP substrates.[26] The photocurrent spectral dependence was acquired by using a tunable light source producing incident light with wavelength dispersion of only ~10 nm of full width at half maximum (FWHM) (TLS120Xe from Bentham Instruments Limited). The incident light wavelength was swept from 600 nm to 800 nm with a 2 nm step. On the other hand, the photoresponse as function of time experiments were acquired using a Fiber-Coupled LED (Thorlabs Inc.) with center wavelength of 625 nm and 15 nm of FWHM with time steps of ~20 ms.[30]

## 3. RESULTS AND DISCUSSION

### 3.1. Micro-reflectance and PL spectroscopy

The first part of this work consists of the analysis through optical micro-reflectance spectroscopy and PL measurements of MoS2 monolayer and bilayer flakes on PP. Figure 1 shows typical micro-reflectance spectra used in this work to compare the influence of the three different encapsulations on the strain tunability of optical properties by applying mechanical uniaxial strain (Figure 1a) or thermal strain (Figure 1b). For the non-encapsulated sample under mechanical strain (Figure 1a-left), it is easy to identify two peaks related to the A and B excitons corresponding to the direct bandgap transition at the K point of the Brillouin zone in monolayer $MoS_2$. The peaks show a mild red shift when 0.9% mechanical strain is applied.

Based on the curve fits applied for each reflectance measured curve, with strain steps of 0.1%, the energy position of the A exciton peak was tracked as function of the strain as shown in Figure 1c. At 1.0% the non-encapsulated sample shows an abrupt variation on the energy shift trend with the strain increase, defined as the maximum achievable strain value for this sample. A linear adjustment fit is applied to each reflectance group of data, as shown on Figure 1c, excluding the initial data points, which often reflect a flake and substrate settling before reaching the observed linear trend. The gauge factor of the sample can be determined as the fit-line slope. A gauge factor of -33 meV/% for the A exciton and -32 meV/% for the B exciton were calculated for this non-encapsulated sample (Figure 1c).

On the other hand, the spectra in Figure 1a-right for the PVFM encapsulated sample shows an inverted profile due to an optical interference effect expected for the case where the sample presents a sub-micron thick and a transparent capping layer with a refractive index that is different than the substrate's.[7] The peaks show a higher red shift when 0.9% mechanical strain is applied.

Using the same curve fit, but with inverted intensities, the spectra can be fitted accurately and a gauge factor of -67 meV/% for the A exciton and -51 meV/% for B the exciton were calculated.

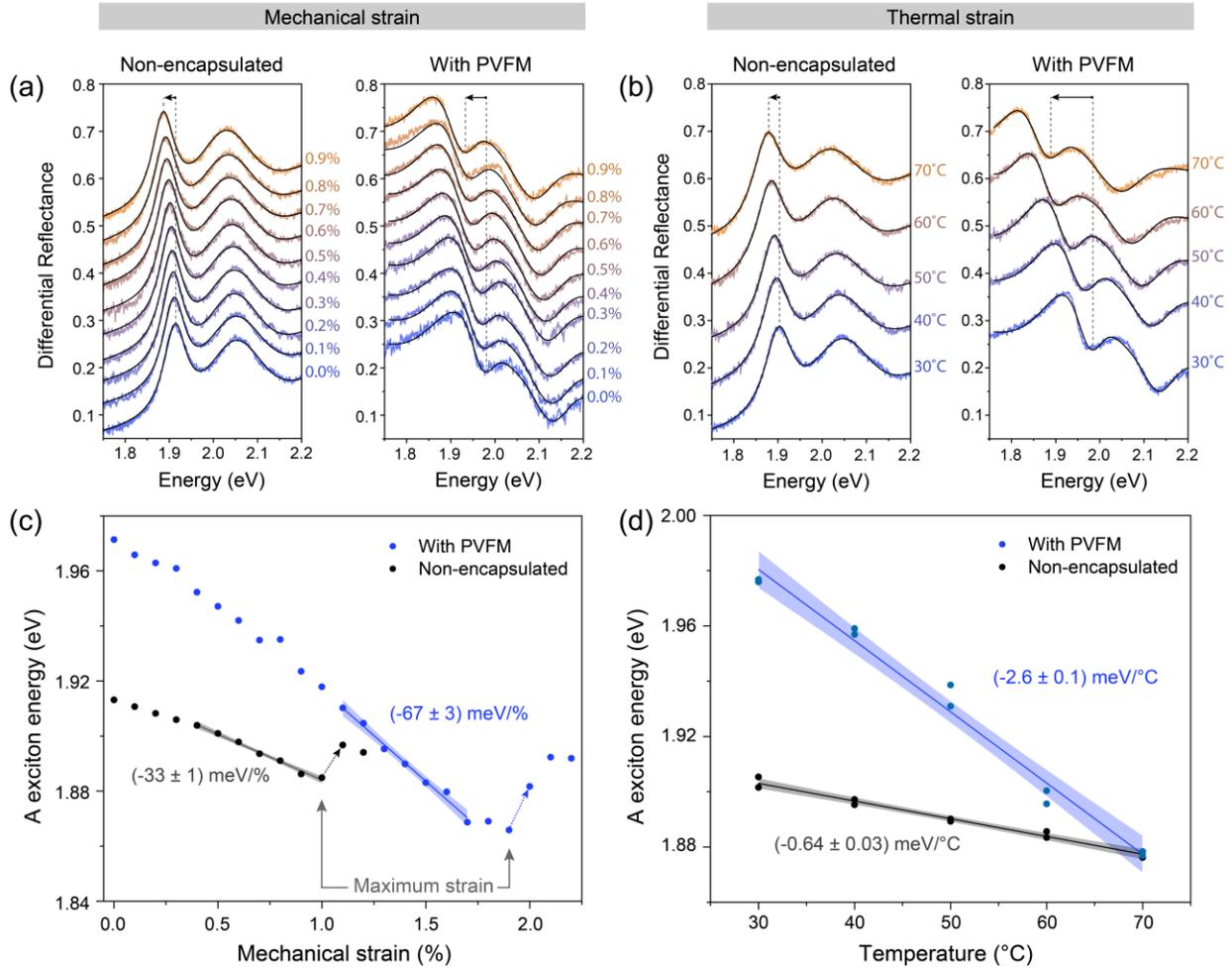

**Figure 1.** (a) Micro-reflectance spectra under mechanical strain ranging from 0% to 0.9% in increments of 0.1%, and (b) micro-reflectance spectra under thermal strain from 30 °C to 70 °C in increments of 10 °C, both for monolayer $MoS_2$ samples without (left) and with PVFM encapsulation (right). For both mechanical and thermal strain, only the initial curves at 0% strain and 30 °C, exhibit unaltered intensity. Intensities of subsequent spectra were vertically shifted for clarity in the stacked plots. Each spectrum was fitted with a curve composed of three Gaussian peaks and one linear function. The energy positions corresponding to the A exciton peak are displayed as a function of mechanical strain in (c) and of thermal strain in (d). Arrows in (c) indicate the maximum achieved strain before a sudden shift in energy position occurs. A linear fit with confidence bands yields slopes of (-33 ± 1) meV/% and (-67 ± 3) meV/% for non-encapsulated (black) and PVFM-encapsulated (blue) samples under mechanical strain, and slopes of (-0.64 ± 0.03) meV/°C and (-2.6 ± 0.1) meV/°C for non-encapsulated (black) and PVFM-encapsulated (blue) samples under thermal strain.

The same procedure was applied to samples under thermally induced strain, resulting in the data presented in Figure 1d. To isolate the spectral shift caused by biaxial strain due to the thermal expansion of the PP substrate, it is necessary to subtract the intrinsic thermal band gap redshift of the sample. In a previous work, the thermal dependence of the energy shift for mono-, and bilayer $MoS_2$ on a $Si/SiO_2$ substrate, which has negligible thermal expansion, was calculated to be -0.44 meV/°C and 0.45 meV/°C respectivelly.[31] By subtracting this value from the slopes calculated in Figure 1d, the thermal gauge factor was determined to be -0.20 meV/°C for non-encapsulated sample (blue) and -2.2 meV/°C for the PVFM-encapsulated sample (black).

The data analysis for all 26 samples can be found in Support Information and the results are summarized in Figure 2, including non-encapsulated samples, as well as samples with PC, PVFM and PMMA polymer encapsulations. The A exciton average gauge factor and the average maximum achievable strain, as well as its standard deviation, are shown for mono- and bilayer samples under mechanical and thermal applied strain (for both A and B exciton see Figure S22 and S24 on Supplementary Material). For the thermal strain results, the gauge factors were calculated already subtracting the intrinsic thermal shift of -0.44 meV/°C and -0.45 meV/°C for mono- and bilayer $MoS_2$.[31]

For the cases of PVFM and PMMA encapsulation, we observe a ~2-fold increase in mechanical strain gauge factor, from around -26 meV/% to around -52 meV/%, and a ~6-fold increase in thermal strain gauge factor, from around -0.20 meV/°C to around -1.45 meV/°C, whereas the increase of gauge factor for PC encapsulation seems to be at the middle way. This is consistent with the fact that PC is the one with lower Young's modulus from the three, ~2.3 GPa,[32] against ~5.5 GPa[25] and ~4.0 GPa[32] for PVFM and PMMA respectively.[32] Also, these results indicate that, in addition to enhancing strain transfer through clamping, as observed under mechanical strain, the thermal expansion strain is significantly influenced by the encapsulant layer. PMMA and PVFM layers exhibit a high thermal expansion coefficient, contributing additional strain to the uppermost region of the sample, beyond that induced by the polymer substrate alone.

The observed average thermally induced strain gauge factors correspond to strains rates of approximately 0.10% per 10°C and 0.31% per 10°C, for the non-encapsulated and PVFM-encapsulated samples, respectively. Considering the ambient temperature of 26°C and a maximum temperature of 70°C used in this work, this temperature range corresponds to around 0.44% and 1.36% mechanical strain, respectively. In regarding to the maximum achievable strain, all polymer encapsulations showed around the same performance, increasing from around 1.4 % to around 2.3 %. This means an energy bandgap change of ~-110 meV for monolayer $MoS_2$ using the average values for the maximum achieved strain and gauge factor for uniaxial mechanical strain for PVMF and PMMA encapsulation. The maximum spectrum energy shift measured was -156 meV for Monolayer $MoS_2$ with PMMA encapsulation (see Supplementary Material Figure S18).

In other studies, the strain gauge factor for non-encapsulated $MoS_2$ monolayer has been reported to range from -3 meV/% to -78 meV/%, with maximum strain values between 0.4% and 1.5%.[5,7,20,33–37] For encapsulated $MoS_2$ samples, the strain gauge factor has been reported to vary more widely, from -42 meV/% to -136 meV/%, with maximum strain ranging from 0.6% to 2.9%.[7,8,22–24,38,39] The highest gauge factor of -136 meV/% was observed in a monolayer $MoS_2$ sample with PVA substrate/encapsulation.[8] However, the same research group later reported a maximum gauge factor of only -63 meV/% using the same method.[40] Notably, a gauge factor of -99.5 meV/% was achieved with $MoS_2$ on a PC substrate and adamantane encapsulation.[7]

To compare our results with those reported in the literature, a fairer comparison is to focus on the improvement achieved with polymer encapsulation rather than the absolute gauge factor, since our study evaluates the effect of encapsulation rather than the entire substrate-sample-encapsulation system. The substrate composition significantly influences strain transfer, which explains why the average gauge factor observed here for the PP substrate, around -26 meV/%, is lower than that reported for PC[7] or polyethylene terephthalate (PET),[33] but higher compared to PDMS.[20] The approximately 2-fold improvement in strain gauge factor observed for PVFM and PMMA in this study is consistent with the results reported in Ref.[7], where a 1.8-fold improvement in the A-exciton gauge factor was observed for monolayer $MoS_2$ on a PC substrate with adamantane encapsulation.

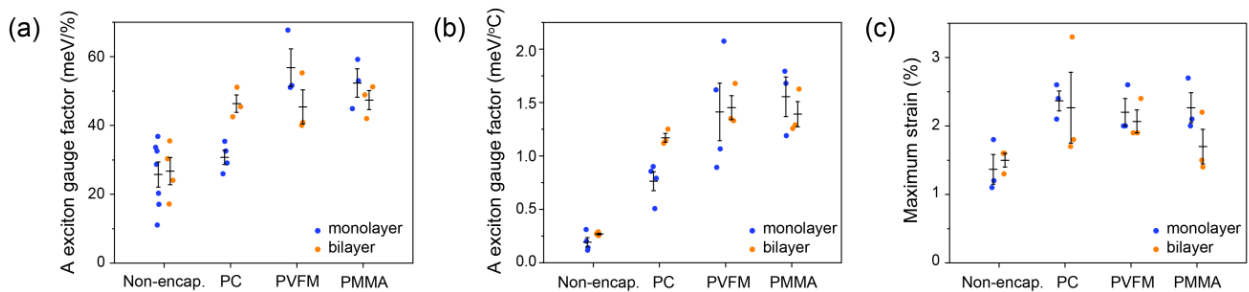

**Figure 2.** Micro-reflectance results of 26 $MoS_2$ samples on PP substrate, including monolayers (blue) and bilayers (orange), under uniaxial mechanical (a,c) and thermal (b) strain, for which the mean values and standard deviations are exhibited. (a,b) the A exciton gauge factor and (c) the maximum achievable strain measured for non-encapsulated samples, and with encapsulation made with PC, PVFM and PMMA polymers.

To see how the strain is spatially transferred to the MoS$_2$ we performed PL hyperspectral images for 0% and for 1% mechanical strain on a monolayer flake encapsulated with PVFM (Figure 3). On Figure 3b, the color of each pixel renders the PL peak energy position as indicated on the color bar scale at the bottom. We observed PL red shifts of ~50 meV and ~30 meV at the middle of the flake and at an area closer to the flake edge, respectively. This change is attributed to the edge effect also reported and discussed on Ref.[29] Although this could affect the result from a single flake depending on the place the optical differential reflectance is measured, the statistical results from 26 samples (each exceeding 40 μm² and measured at least 3 μm from the edge) and the repeatability of the method used minimizes any edge effect contribution on the average results.

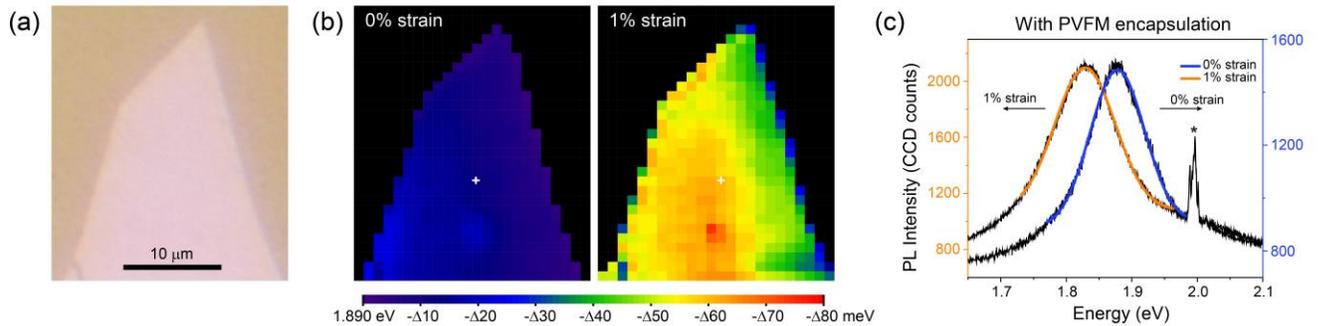

**Figure 3.** (a) optical microscopy image of a monolayer MoS$_2$ sample on PP substrate and encapsulated with PVFM used for the PL intensity color hyperspectral images for 0% and 1% mechanical strain. (b) The heat-color palette renders the PL peak position from 1.890 eV (dark blue) to 1.810 eV (red). (c) Spectra acquired at the position of the white cross markers on PL maps for 0% (blue) and 1% (orange) mechanical strain. A single gaussian was used for curve fitting the PL peak.

## 3.2. Photocurrent

The second part of this study investigates the opto-electronic photocurrent response of photosensor devices composed of a bilayer $MoS_2$ flake on the top of Au/Ti electrodes on PP substrates. To explore this, a few photodetector devices were fabricated, both with and without PVFM encapsulation, to serve as a case study of the strain-tuned opto-electronic photocurrent response (Figure 4). For real-world applications, it is crucial that the 2D semiconductor used in the photosensor of FET devices exhibits a fast response, with the photocurrent increasing linearly with light intensity and maintaining good temporal stability. Equally important is the reproducibility of its response, which ensures reliable device performance. This means working in the photoconductive regime, where the minority charge carrier traps don't play a significant role. Otherwise, artefacts such as the cumulative effect of previous exposures or even signal saturation could affect the accuracy of the photocurrent reading when the device is exposed to higher intensity light pulses.

Interestingly, as we observed on Figure 4a for the photocurrent response of a bilayers $MoS_2$-based photodetector subjected to pulses of increasing light power, the photocurrent response is increased, and the photocurrent noise and fluctuation are reduced when the same device is encapsulated with PVFM. In addition, a faster response time is also observed upon encapsulation PVFM encapsulation. These results seems to be attributed to an n-type doping of the PVFM polymer, in a similar way observed for PVP[18] and HfO[41] encapsulations. The oxygen atoms in the PVFM structure may contribute to electron donation to the semiconductor, thereby reducing charge trapping by removing adsorbates such as $O_2$ and $H_2O$ molecules.[18] This leads to a reduction of the photogating effect, making the device respond faster and more reproducible way. However, further investigation is needed to confirm the exact mechanism behind this effect. In Figure 4b, the photocurrent is analyzed as a function of the incident light power density. For the device after encapsulation, the photocurrent response goes linear with power even for 300 mW/cm$^2$, which is almost three times higher incident power density when compared to the non-encapsulated device that loses the linearity with light power after 100 mW/cm$^2$.

To directly measure the energy band shift for this type of device with and without PVFM encapsulation, we analyzed the responsivity as function of the incident light energy. In Figures 4d and 4e, we observed an optical response energy shift of 8 meV (16 meV) when the non-encapsulated (PVFM encapsulated) device is subject to a 0.4% mechanical strain. This means an improvement of gauge factor from ~-20 meV/% to ~-40 meV/% after encapsulation. Notably, these values are comparatively lower than those derived from optical measurements, primarily due to the limited strain range (0-0.4%) used in this analysis, which introduces increased uncertainty. Nevertheless, they still reflect a twofold improvement as previously reported.

In Figure 4f, the effect of encapsulation on device lifetime is examined by measuring the device responsivity over time when exposed to ambient conditions (22 °C and 45% RH). For this analysis, two bilayer MoS$_2$-based devices were fabricated: one without any capping layer and one with PVFM encapsulation. The responsivity of the non-encapsulated device (black) dropped to 50% on the 9[th] day, whereas the PVFM-encapsulated device (blue) maintained 50% responsivity until the 28[th] day. Although these results are based on only two devices, they suggest that PVFM encapsulation can improve device lifetime as it prevents intrusion of ambient molecules such as O$_2$ and H$_2$O into the material, thus reducing charge carrier traps.

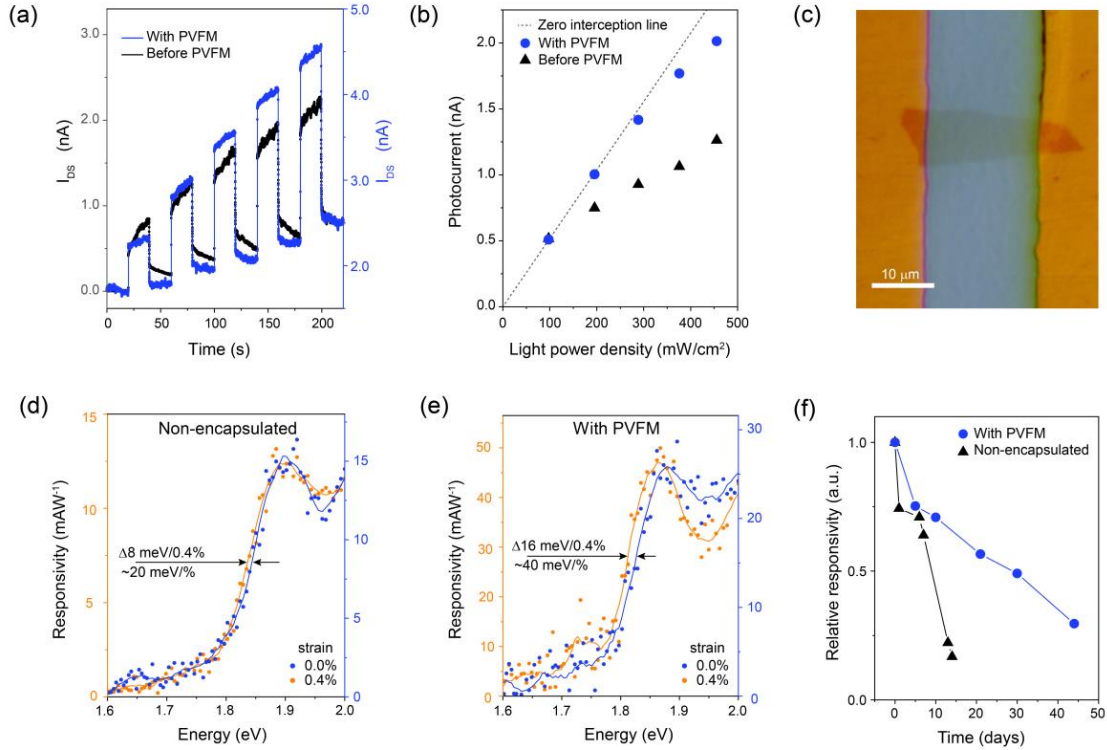

**Figure 4.** (a) Optical responsivity of a MoS$_2$ bilayer-based device on PP substrate, shown on the optical image in (c), before (black) and after (blue) PVFM encapsulation. The device was homogeneously exposed to 20 seconds long incident light pulses with varying powers densities of 97, 195, 289, 376, and 456 mW.cm$^{-2}$ ($\lambda$ = 625 nm and 15 nm of FWHM). (b) photocurrent measured for the device before (black) and after (blue) PVFM encapsulation as function of the light power density. The dashed grey line is a zero-interception line [Y(x) = 0.0051.x] to show the linearity of the responses. (d,e) spectral dependence of the photoresponse for two devices composed of MoS$_2$ bilayers on top of PP substrates and Au/Ti electrodes, without and with PVFM encapsulation, for 0% (blue) and 0.4% (orange) mechanical strain. The continuous lines represent a smooth interpolation using a polynomial smooth interpolation using polynomial of 2$^{nd}$ order to help on the energy shift estimation. An energy shift of -8 meV and -16 meV was observed, in the devices without and with PVFM encapsulation, respectively. (f) comparison of device lifetime in ambient conditions for non-encapsulated (black) and encapsulated (blue) devices on PP substrate.

## 4. CONCLUSIONS

In summary, we have carried out a systematic study of the improvement on the transferable strain on $MoS_2$ monolayer and bilayer flakes on flexible PP substrates when three distinct and simple to perform polymer encapsulations are applied. We found that PVFM and PMMA encapsulation can increase the mechanical strain gauge factor by about 2-fold and the thermal strain gauge factor by about 6-fold, whereas for PC encapsulation the strain gauge factor enhancement was slightly lower. The same 2-fold increase in mechanical strain was also observed for $MoS_2$ bilayer-based devices on PP substrates for PVFM encapsulation. The three polymers have also increased the maximum achieved strain from about 1.4% to about 2.3% strain. As a case for comparative analysis of the optoelectronic response without and with encapsulation, we have fabricated and measured $MoS_2$ bilayer-based devices on PP substrates. We observed an improvement of the device quality when encapsulated with PVFM, from which we can highlight the increase of the photocurrent and device response time, as well as a reduction of the photocurrent fluctuations. Additionally, we demonstrated an increase in device lifetime upon PVFM encapsulation.

This work based on the analysis of 26 $MoS_2$ samples indicates that by a simple method of spin-coating PMMA or PVFM polymer solutions to encapsulate $MoS_2$ samples on a flexible substrate, the strain transfer can be highly enhanced while at the same time increasing the devices' lifetime and optical response quality as it protects the sample from ambient absorbents. The photosensor results with PVFM encapsulation are very promising and can have a positive impact on the straintronics industry as this polymer is inexpensive and widely available.

ASSOCIATED CONTENT

Supplementary materials include AFM and optical images of the polymer layers used to measure thickness, optical microscope images of $MoS_2$ samples before and after PVFM encapsulation, and optical differential micro-reflectance data for A and B excitons under mechanical and thermal strain across all 26 samples used in Figures 1 and 2.

AUTHOR INFORMATION

**Corresponding Authors**

**Andres Castellanos-Gomez** − 2D Foundry research group, Instituto de Ciencia de Materiales de Madrid (ICMM-CSIC), E-28049 Madrid, Spain; https://orcid.org/0000-0002-3384-3405; Email: andres.castellanos@csic.es

**Thiago L. Vasconcelos** – Materials Metrology Division, Instituto Nacional de Metrologia Qualidade e Tecnologia (INMETRO), 25250-020 Duque de Caxias, RJ, Brazil; 2D Foundry research group, Instituto de Ciencia de Materiales de Madrid (ICMM-CSIC), E-28049 Madrid, Spain; https://orcid.org/0000-0003-0195-444X; Email: tlvasconcelos@inmetro.gov.br

**Authors**

**Simeon N. Vladimirov** − 2D Foundry research group, Instituto de Ciencia de Materiales de Madrid (ICMM-CSIC), E-28049 Madrid, Spain.

**Onur Çakıroğlu** − 2D Foundry research group, Instituto de Ciencia de Materiales de Madrid (ICMM-CSIC), E-28049 Madrid, Spain.

**Carmen Munuera** – 2D Foundry research group, Instituto de Ciencia de Materiales de Madrid (ICMM-CSIC), E-28049 Madrid, Spain; https://orcid.org/0000-0001-8524-9304

ACKNOWLEDGMENTS

*The Ministry of Science and Innovation (Spain) through Projects PDC2023-145920-I00, TED2021-132267B-I00 and PID2020-115566RB-I00. The authors also acknowledge funding from the Comunidad de Madrid through the CAIRO-CM Project (Y2020/NMT-6661) and European Union's Horizon 2020 research and innovation program under the grant agreement 956813 (2Exciting).*